\documentclass[%
 aip,
 jmp,%
 amsmath,amssymb,
 reprint,%
]{revtex4-1}

\usepackage{graphics}
\usepackage{graphicx}
\usepackage{dcolumn}
\usepackage{bm}

\begin{document}

\preprint{AIP/123-QED}

\title{Magnetoplasmon excitations at graphene vortex Hall fluid edge}
\thanks{Corresponding author. Email: mrabiu@uds.edu.gh.}

\author{M. Rabiu}
 \email{mrabiu@uds.edu.gh}
 \affiliation{Department of Applied Physics, Faculty of Applied Sciences, University for Development Studies, Navrongo Campus, Ghana.
}

\author{S. Y. Mensah}%
\affiliation{Department of Physics, Laser and Fiber Optics Center, University of Cape Coast, Cape Coast, Ghana.
}%

\author{S. Y. Ibrahim}
 \affiliation{University for Development Studies, Faculty of Mathematical Science, Department of Mathematics, Navrongo, Ghana.
}

\author{S. S. Abukari}
\affiliation{Department of Physics, Laser and Fiber Optics Center, University of Cape Coast, Cape Coast, Ghana.
}%

\date{\today}

\begin{abstract}
We investigate magnetoplasmon dynamics localized on the edges of graphene vortex Hall fluid. The vortex matter captures an anomalous term that causes vortex localization near fluid boundary and creates a double boundary layer, $\Delta_0\propto (\beta-1)\ell_B$ with $\beta$ being filling factor. The term also has qualitative effect on resonant excitations of edge magnetoplasmons. We found that for sharp edges under experimental conditions, graphene Edge Magnetoplasmon (EMP) resonances have similar behavior as in recent experiments. Gradual distinctions arise for smooth edges in the presence of the anomalous term, where a weak EMP peak appear. The second peak becomes well noticed as the smoothness is increased. We identified the resonant mode as an Inter-EMP. It originates from the oscillations of charges in the inner boundary of the double layer. The present observation brings to light the direct cause of Inter-EMP which remained to be detected in graphene experiments.
\end{abstract}

\pacs{}
\keywords{Euler hydrodynamics, Quantum hall fluid, Vortex dynamics, Fractional quantum hall state, Magnetoplasmons.}
\maketitle

\section{Introduction\label{Sec:Section1}}
Quasi-particle excitations of collective oscillations and rotations of charges coupled to magnetic field, edge Magnetoplasmons, are gapless modes localized on perimeter of pool of electron gas. In conventional two-dimensional electron gas systems (2DEGS), magnetoplasmon frequency, for a bounded electron liquid, splits into bulk and edge modes. The edge magnetoplasmons propagate unidirectionally along the perimeter with chirality determined by the sign of an applied magnetic field. Graphene has emerged as a new kind of 2DEGS which is a two-dimensional material of carbon atoms arranged in honeycomb lattice. Magnetoplasmon frequencies in Graphene are unique and observed at rather low excitation energy. The behavior can be attributed to the ultra-relativistic nature of charges combined with its linear band structure. 

In general, finite samples are characterized by physical edges. Edge states come from one-body confinement potential from the boundary which bends energy levels up or down near the edge. The resultant effect is that charges are pushed over a chemical potential sandwiched between two Landau energy levels. The source of the confinement potential is very different between 2DEGs and graphene. It is provided by dopants in the 2DEGS and solely by presence of graphene boundaries.

The study of edge charge dynamics in quantum Hall regime has gained renewed interest both theoretically ~\cite{OGBalev2011, WeihuaWang2011,WeihuaWang2011b} and experimentally~\cite{NKumada2014, IvanaPetkovic2013, IPetkovic2012}. In particular, edge Magnetoplasmons (EMPs) in graphene is studied only recently , over a decade after its successful isolation in 2004.  However, previous studies only investigated EMPs at $\nu =2$ Integer Quantum Hall Effect (IQHE) and based on sharp electron density profiles at the boundary. Even at this state,  EMP dynamics in graphene are already different from the 2DEGs. For large system size that is best suited for hydrodynamic description, radio-frequency EMP resonance peaks are detected on Hall plateaus~\cite{Wassemeir1990} in 2DEGs.  In ref.~\cite{WeihuaWang2011b}, magnetoplsmon resonances in non-Quantum Hall regime was observed. These resonant excitations are yet to be investigated in graphene within fractional quantum Hall regime. To explore further, we employed a quantized point vortex hydrodynamics theory of Euler equations. Quantized vortices (QVs) have been widely observed in physical systems; Liquid helium, type-II superconductors and ultracold atom gases in optical lattices. Fractional Quantum Hall Effect (FQHE) is another example. In particular, upon application of perpendicular magnetic field to Dirac fluid in graphene, vortices arise from motion of Dirac fermions near K and K$^\prime$ valleys~\cite{BitanRoy2012}. The magnetic fluxes are viewed as topological defects capable of deforming fluid boundaries. In the limit of large vortices-caused by strong magnetic field, the vortex matter alone fully characterized the flow and dynamics of electrons is properly described by equations of classical hydrodynamics. Thus, the use of microscopic electronic theory becomes a matter of choice. Like electrons, in finite incompressible fluid vortices  are pushed to sample boundary and are drift along it by a potential gradient leading to an excitable one-dimensional system. 

We are wondering to investigate basic features of low-energy magnetoplasmon excitations localized and propagating at edges of quantum vortex Hall fluid, using hydrodynamic Euler and continuity equations. We observed EMP excitations on fractional vortex Hall plateaus. The rest of the paper is organized as follows. In Sect.~\ref{Sec:Section2}, we present system model hydrodynamical equations with velocity solution that captures edge dynamics in a disk geometry. The boundary of the fluid is also modeled to reflect its smoothness. Boundary frequency oscillations are derived for a generic Hall fluid edge. A discussion of the main results is finally presented in Sec.~\ref{Sec:Section3}, where we draw a conclusion and make some recommendations.

\section{\label{sec:The-Theory} Theoretical model\label{Sec:Section2}}
\subsection{Hydrodynamical equations and point vortices}
To begin our analysis of magnetoplasmon excitations at the edge of Hall fluid, let us consider an idealized, isentropic, inviscid fluid. The Classical governing equation for a Dirac electron fluid are the Euler and continuity equations;
\begin{equation}
 D_t\rho_{\alpha} = 0\quad \mbox{ and }\quad D_t{\bf u}_{\alpha} + \nabla {\rm p}_{\alpha}= 0,\label{Eq:EMFluid}
\end{equation}
respectively. Where $D_t \equiv \partial_t + {\bf u}\cdot {\bf \nabla}$ and ${\bf u}$ is macroscopic fluid velocity connected with the microscopic electron velocity, ${\bf v} = v_F{\bf k}/k$. ${\rm p}=en_o V$ is the partial pressure which modifies momentum flow at the boundary. $n_0$ is electronic fluid density and $V$ is confinement potential. The fluid component index $\alpha\equiv$ (K$^\prime\uparrow$, K$\uparrow$, K$^\prime\downarrow$, K$\downarrow$). ${\rm K, K^\prime}$ and $\uparrow,\downarrow$ are the valley and spin indexes, respectively. Taking the curl of Eq. (\ref{Eq:EMFluid}), we get the Helmholtz equation
\begin{equation}
  D_t\omega_{\alpha} = 0.\label{Eq:MVFluid}
\end{equation}
Where the quantity $ \omega= {\bf \nabla}\times {\bf u}$ is vorticity which we assumed to be non-zero for incompressible flows. $\omega$ is related to magnetic fluxes. It is considered as frozen into the fluid and constitute its own fluid. The continuity equation for the vortex fluid $ D_t\rho_{{\rm v},\alpha} = 0$, also holds. We regard vortices as localized solutions of the Euler equation. In two-dimensional disk geometry, a point vortex solution exists. The Kirchoff equation obtained from Eq.~\ref{Eq:MVFluid} for the motion of fluid particles is~\cite{Flutcher1999}
\begin{equation}
  {\rm u}_{\alpha} =  {\rm i}\sum_{i=1}^{N_{\rm v}}\frac{\Gamma_{i,\alpha}}{z_{\alpha} - z_{i,\alpha} (t)} - {\rm i}\sum_{i=1}^{N_{\rm v}}\frac{\Gamma_{i',\alpha}}{z_{\alpha} - z_{i',\alpha}(t)}. \label{eq:MFluidvelo}
\end{equation}
Where $z = x + {\rm i} y$, $u = u_x - {\rm i} u_y$, $\Gamma$ is the circulation and $\Omega$ which is identified with cyclotron frequency of the fluid particles. The edge effect is introduced by image vortices located at $z_{i'}$. Assuming a flow in which the circulation, $\Gamma_{i,\alpha}$ ($=\Gamma_{\alpha}$) is both minimal and chiral so that in the limit, $N_{\rm v} \to \infty$ rotation is compensated by the large number of vortices. For magnetic field, $B\neq 0$, vortices of all components are smoothly distributed with fixed mean density obtained from Faymann formular, $\rho_0 = (\lambda/\pi)(\Omega/\Gamma_{\alpha})$. One also has Bohr-Sommerfeld phase-space quantization of the circulation $m_{\text{v}}\Gamma_{\alpha} = \beta_{\alpha}\hbar$. $\lambda$ is degeneracy of the system and $m_{\text{v}}$ is inertia of vortex flow with  an integer $\beta_{\alpha}$. The equation for the slow motion of  vortices is easily obtained from Eq.~\ref{eq:MFluidvelo}. It is given by~\cite{Wiegmann2013}
\begin{equation}
  {\rm v}_{i,\alpha}(t) = {\rm  i}\sum_{j\neq i}^{N_{\rm v}}\frac{\Gamma_{\alpha}}{z_{i,\alpha}(t) - z_{j,\alpha}(t)} 
-{\rm  i}\sum_{j\neq i}^{N_{\rm v}}\frac{\Gamma_{\alpha}}{z_{i, \alpha}(t) - z_{j',\alpha}(t)}. \label{Eq:Vortex}
\end{equation}

In the following, we will specialize in the zero energy state of the system where K and K$^\prime$ components of the fluid decouple. Dynamics are then localized in either sublattice and $\sum_{\alpha}{\rm v}_{i,\alpha}=\lambda {\rm v}_i$.

\subsection{Velocity on the edge of quantized vortex Hall fluid} 
Writing {\rm v} interms of {\rm u} using the identities $\pi\delta(r) \equiv \bar{\partial }(1/z)$ and   $\sum_{i\neq j}[2/(z-z_i)(z_i-z_j)] \equiv [\sum(1/(z-z_i)]^2 - \sum[1/(z-z_i)]^2$ , it is straight forward to show that~\cite{Wiegmann2013} $  {\rm v} = {\rm u}+({\rm i}\Gamma/2)\partial_z{\rm log}\rho_{\rm v}$. We can recast the expression as
\begin{equation}
  \rho_{\rm v}{\rm v} = \rho_{\rm v}{\rm u}+\frac{1}{2}\Delta{\rm u} + \rho_{\rm v} {\rm v}_D. \label{Eq:EFlux}
\end{equation}
Where $v_D$ ($=e\ell_B^2\nabla V/\hbar$) is the drift velocity of vortices confined by the potential, $V$. In order to capture quantum effects needed to qualitatively explain experimental results, we sort a quantized version of Eq.~\ref{Eq:EFlux}. A quantization rule is imposed on the hydrodynamic variables u and $\rho_{\rm v}$. i.e, $\left[{\rm u},\rho_{\rm v}\right] = -(\hbar/m_{{\rm v}}\Gamma)\Delta u$. Direct substitution in Eq.~\ref{Eq:EFlux} leads to the velocity
\begin{equation}
  {\rm v}(r) = {\rm u}(r)+\frac{1}{2\rho_0}\left(1-\frac{\hbar}{m_{{\rm v}}\Gamma}\right)\Delta{\rm u}(r) + {\rm  }v_D(r), \label{Eq:EFluxQuan}
\end{equation}
which now expressed in polar coordinates. We calculate the value of ${\rm v}$ at the disk edge by integrating out normal modes in the above equation. On the disk boundary with radius $R$, the angular component of the velocity reads
\begin{eqnarray}
  {\rm v}(R) &=& \left({\rm u}(r)+\frac{1}{2\rho_0}\left(1-\frac{\hbar}{m_{{\rm v}}\Gamma}\right)\nabla_n{\rm u}(r)\right)\delta(R-r)\nonumber\\
		&& +  {\rm v}_D(r)\Theta(R-r). \label{Eq:EFluxQuan}
\end{eqnarray}
Where $\nabla_n$ is normal derivative. The first two terms are always connected with the bulk but the last term is due to the boundedness, hence the inclusion of step function. Following standard elementary calculations, the edge velocity ${\rm u(r)}$ computed from Eq.~\ref{eq:MFluidvelo}, on the boundary, becomes
\begin{equation}
	{\rm u}(R) = \sum_i\frac{\lambda\Gamma_i}{2\pi}\frac{R^2-r_i^2}{|{\bf R}-{\bf r_i}|^2},\quad\text{with}\quad {\bf R} = {\bf r}R/r. \label{Eq:EFluxQuan2}
\end{equation}

Carrying out Fourier transform of Eq.~\ref{Eq:EFluxQuan} using Eq.~\ref{Eq:EFluxQuan2} and following several steps of simplifications, we find that
\begin{eqnarray}
  {\rm v}(QR) &=&\sum_{i,\ell} \frac{\epsilon_{\ell} \lambda\Gamma}{R}\left[\left(\frac{r_i}{R}\right)^{|\ell|}-\frac{Q}{\rho_0R}\left(1-\frac{1}{\beta}\right)\left(\frac{r_i}{R}\right)^{|\ell-1|} \right]\nonumber\\
		&&\times K_{\ell}(QR) +  {\rm v}_D\delta_{\ell,1}J_{\ell}(QR). \label{Eq:EFluxQuan3}
\end{eqnarray}
$Q$ is the wave vector of excitations along the edge. Close inspection of the preceding equation points to some remarkable characteristics of the edge dynamics. In particular, if $r_i < R$, a smooth and sharp maximum peaks can be observed from the second and third terms.

Now, the frequency of edge excitations is calculated from a continuity equation defined on the boundary as $\omega_{EMP} \rho_{e} = Q(\rho_e + \rho_b ){\rm v}$ and in the bulk, $\omega_{ EMP} \rho_{b} = - {\rm i}\nabla[(\rho_e + \rho_b) {\rm v}]$. Where $\rho_b$ and $\rho_e$ are the bulk and edge densities. Putting these two equations together we get an implicit dispersion relation
\begin{eqnarray}
	\omega_{EMP} = Q\,{\rm v}(QR) - {\rm i}\nabla {\rm v}(QR) -{\rm i}{\rm v}(QR)\,\nabla {\rm log}\rho_{\rm v}.    \label{Eq:EMPLs}
\end{eqnarray}
The first term comes from the edge dynamics and the second is a contribution from the the bulk. For $\rho_{\rm v} = \rho_0 + \delta \rho_{\rm v}$ with $\rho_0 >>  \delta \rho_{\rm v}$,  the third term can be neglected. 

It is important to mention that a quantized Equation~\ref{Eq:Vortex} in the complex coordinates ($z,\bar{z}$), immediately leads to Halperin-like fractional quantum Hall states for graphene. We do not wish to extend our arguments in this direction but only to point out the connection of our theory to FQHE.

\subsection{Model of sharp and smooth boundaries}
To model graphene vortex Hall edges, we consider a sufficiently large but finite droplet consisting of $N$ vortices in $N_{orb}$ orbitals occupying a disk with edges at equilibrium radius $R_0 = \ell_B\sqrt{N/\pi\rho_0}$. Theoretically, quantum Hall fluid edges are modeled by sharp boundaries which often lead to an over simplifications and may fail to capture important physics at the edge. Nonetheless, a fundamental EMP mode is always present. In reality, the edges of Hall fluid are never sharp but (smooth and) rounded on the scale of $\Delta_0 \propto \beta(1-1/\beta)\ell_B$. The smoothness is caused by clustering of vortices at the edges and subsequently performing small cyclotron oscillations with period far less than that of the EMPs. It is possible to have dipole moment within the boundary layer $\Delta_0$ which has been largely ignored in literature~\cite{OGBalev2011,WeihuaWang2011}. Modulation of charges in the layer can be related to the change in local radius, $\delta R$. For sufficiently smooth edges this can be expanded in Fourier series as $\delta R(\varphi) = R_0\sum_{m}b_mExp({\rm i}m\varphi) + c.c$ for all possible angular momenta excitations. $m$  is an integer. Where $0 < b_m < 1$ is a geometric parameter. It measures the smoothness of the edge. After some algebra and simplifications, the velocity Equation~\ref{Eq:EFluxQuan3} now takes the form
\begin{eqnarray}
  {\rm v}(QR_0) &=&\sum_{i,\ell,n} \frac{\lambda\beta \hbar}{2\pi m_{{\rm v}}R_0}\Big[\left(\frac{r_i}{R_0}\right)^{|\ell|}F_{1,n}(b_M) - \frac{Q}{\rho_0R_0}\times\nonumber\\
		&&\left(1-\frac{1}{\beta}\right)\left(\frac{r_i}{R_0}\right)^{|\ell-1|}F_{2,n}(b_M)\Big]K_{\ell+n}(QR_0)\nonumber\\
		&&  + {\rm v}_D\delta_{\ell,1}F_{3,n}(b_m)J_{\ell+n}(QR_0). \label{Eq:EFluxQuan4}
\end{eqnarray}
Where we have used the modified Bessel function relation $K_{\ell}(\Lambda x) = \Lambda^{-\ell}\sum_n[(\Lambda^2-1)x/2]^nK_{\ell+n}(x)$ and  the Bessel function $J_{\ell}(\Lambda x) = \Lambda^{-\ell}\sum_n[(1-\Lambda^2)x/2]^nJ_{\ell+n}(x)$ with $\Lambda = \sum_{}b_mExp({\rm i}m\varphi) + c.c$. In Eq.~\ref{Eq:EFluxQuan4}, we have integrated over $\varphi$ to get $F_{1,n}(b_m) = \int d\varphi (\Lambda^2+1)^n$, $F_{2,n}(b_m) = \int d\varphi \Lambda(\Lambda^2 + 1)^n$ and $F_{3,n}(b_m) = (-1)^nF_{2,n}(b_m) $. Tit is interesting to recover the standard sharp edge expression of EMP dispersion used in~\cite{IPetkovic2012, IvanaPetkovic2013}. We carry out a replacement $R\to R_0$ in Eq.~\ref{Eq:EFluxQuan3} (or $n= 0$ and $\Lambda= 1$ in Eq.~\ref{Eq:EFluxQuan4}) which gives $F_{i,0} = 1$, for $i = 1,2,3$.

\section{Results, Discussions and Conclusions\label{Sec:Section3}}
To observe the behavior of $\omega_{EMP}$, we make contact with the edge excitation parameters through the following identification;
\begin{equation}
	QR_0 = \sum_{\ell}\ell\,\delta_{\ell,\Delta L} ,
\end{equation}
where $\Delta L = L-L_0$, is the change in edge angular momentum. We exploit this correspondence to observe edge magetoplasmon excitations. The number of distinct ways which occurs in a particular energy channel  $\omega_{EMP}\{n_{\ell}\}$ having filling fraction, $\beta$ is computed from $W=(\Delta L + \beta -1)!/(\Delta L !\, (\beta-1)!)$. Excitations are weighed by the factor, $W$. A higher $W$ means broader resonance and vice-versa.

\begin{figure}[th!]
	\centering
	\includegraphics[scale=.5]{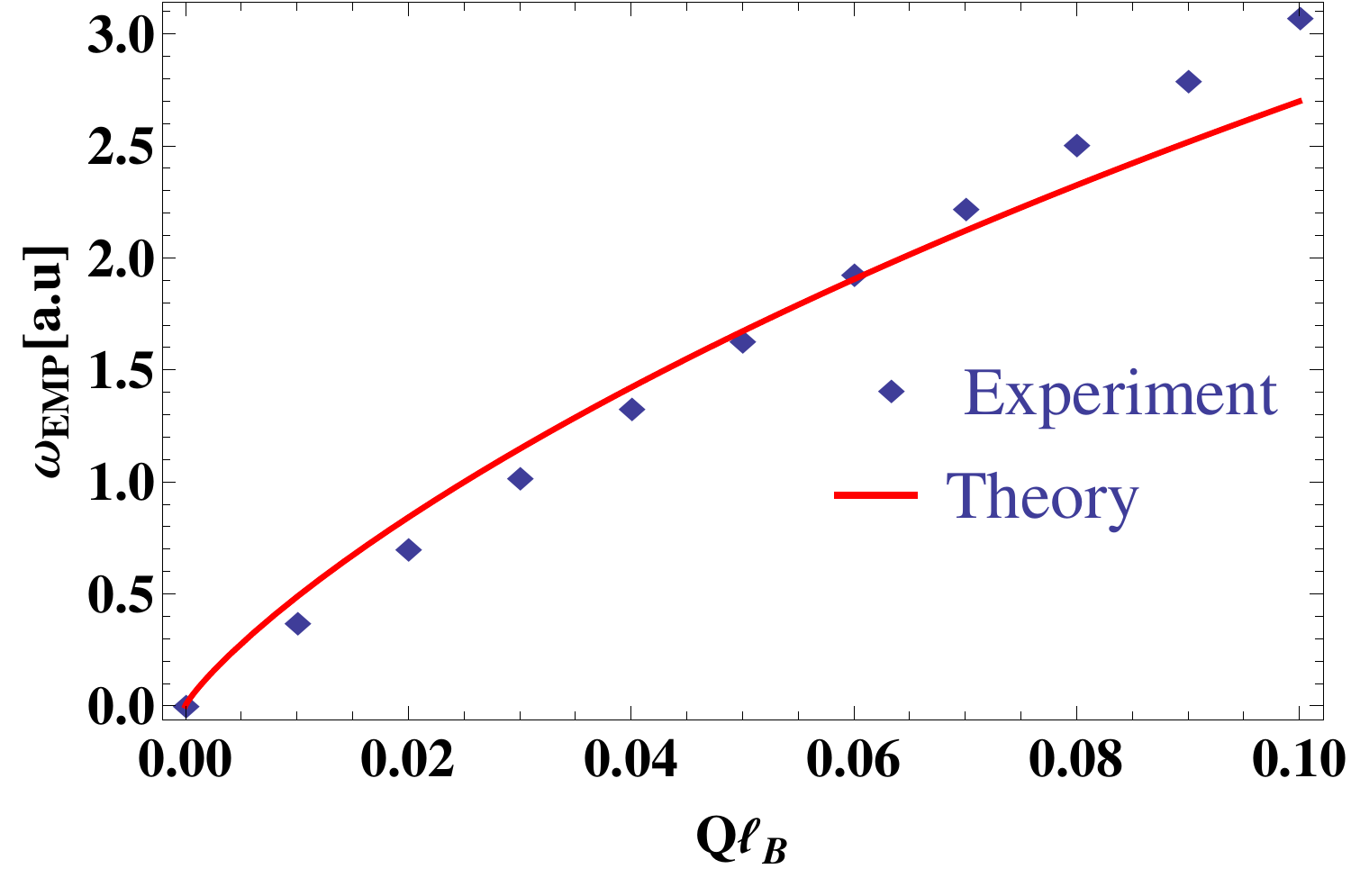}
	\caption{Edge magnetoplasmons dispersion under experimental relevant conditions. Theory is compared with experiment for $\beta = 3$, $N = 6$ and $B \sim 36$ T.}\label{Fig:EMP1}
\end{figure}

\begin{figure*}[th!]
	\centering
	\includegraphics[scale=.5]{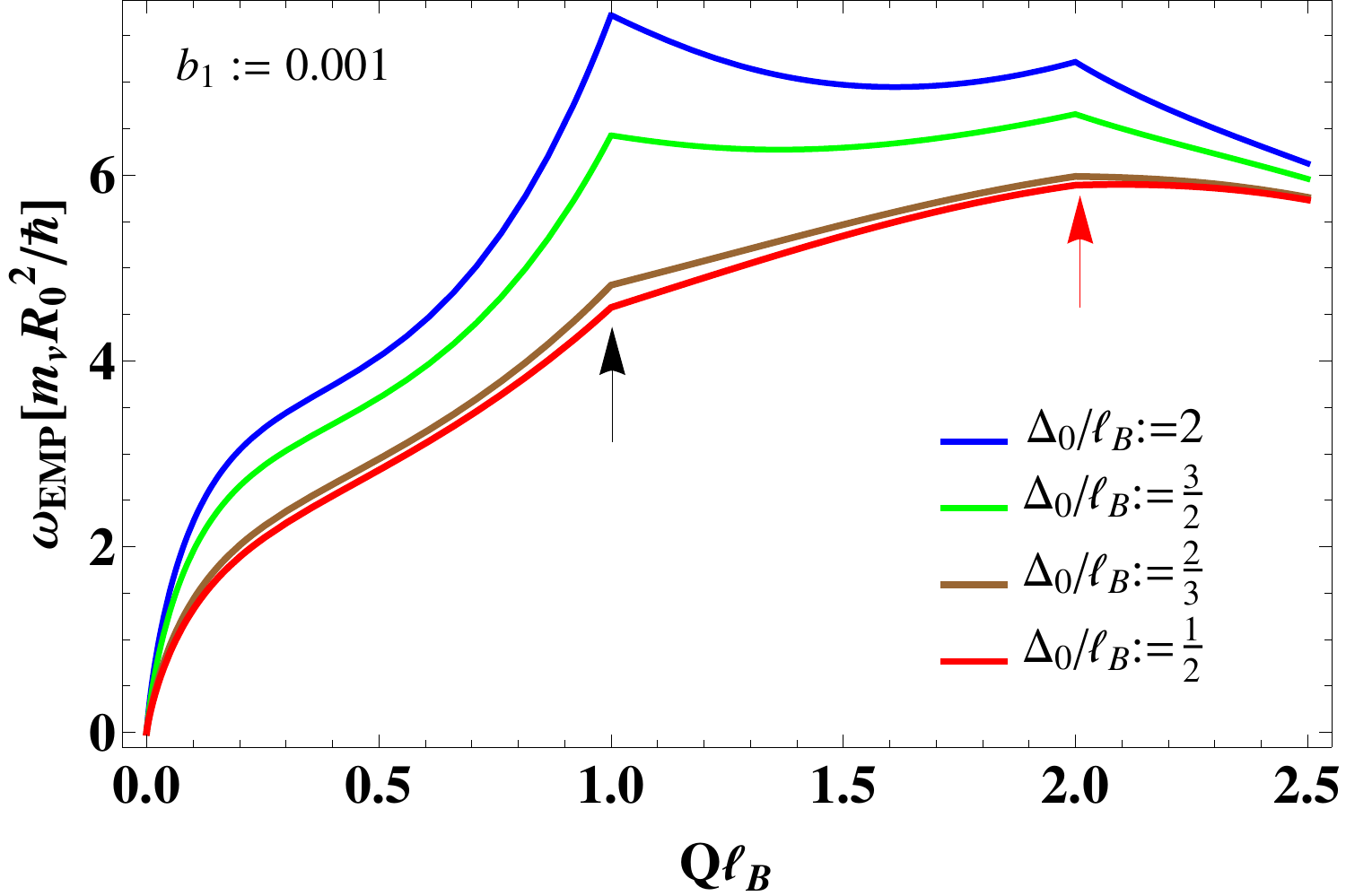}
	\includegraphics[scale=.5]{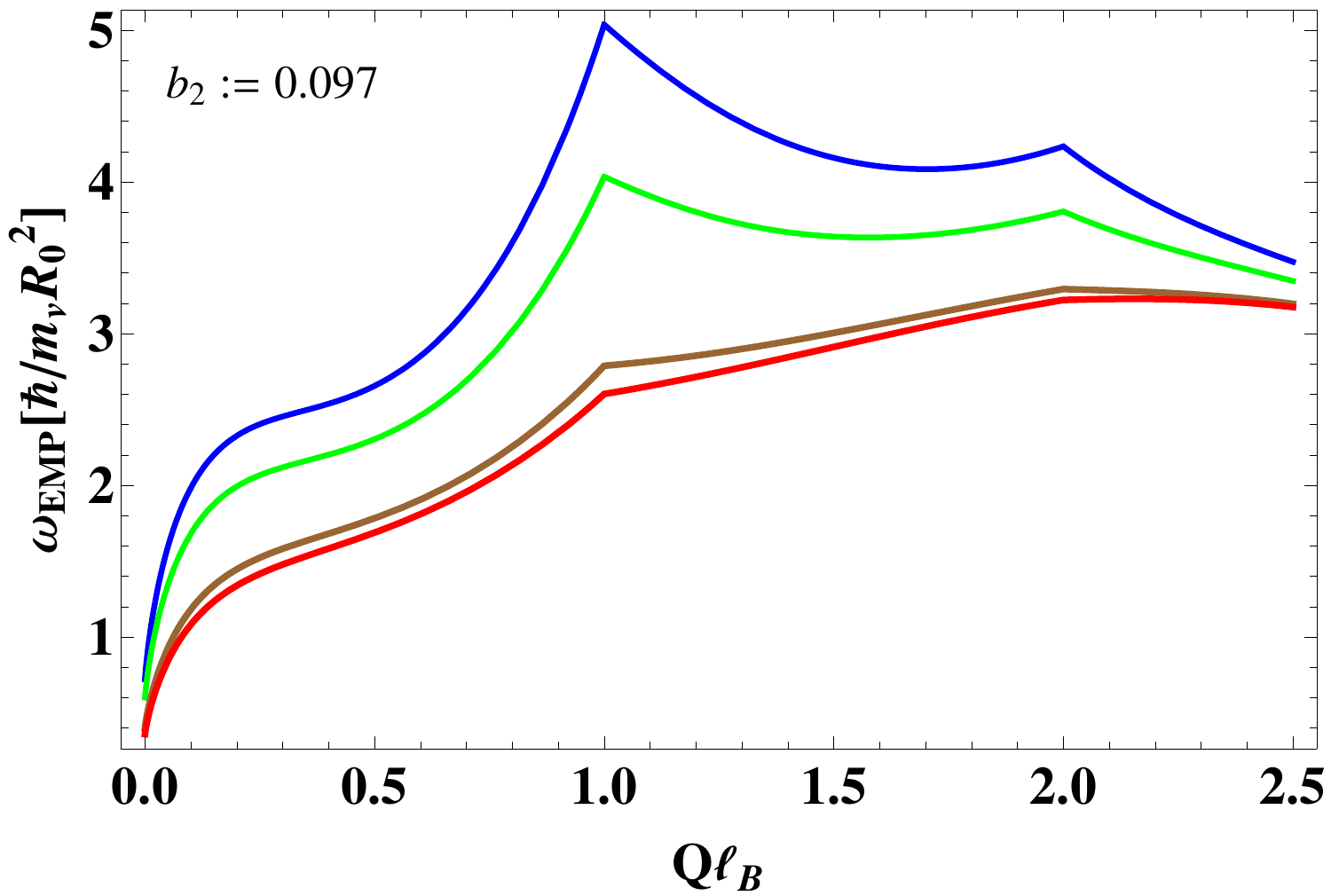}
	\caption{Edge magnetoplasmon dispersion for sharp and smooth edges. The anomalous factor, $\Delta_0/\ell_B$ is varied from $2.0$ to $ 0.5$. Geometry factor: (Left) $b_1=0.01$ and (Right) $b_2=0.097$.}\label{Fig:EMP3}
\end{figure*}

\begin{figure*}[th!]
	\centering
	\includegraphics[scale=.5]{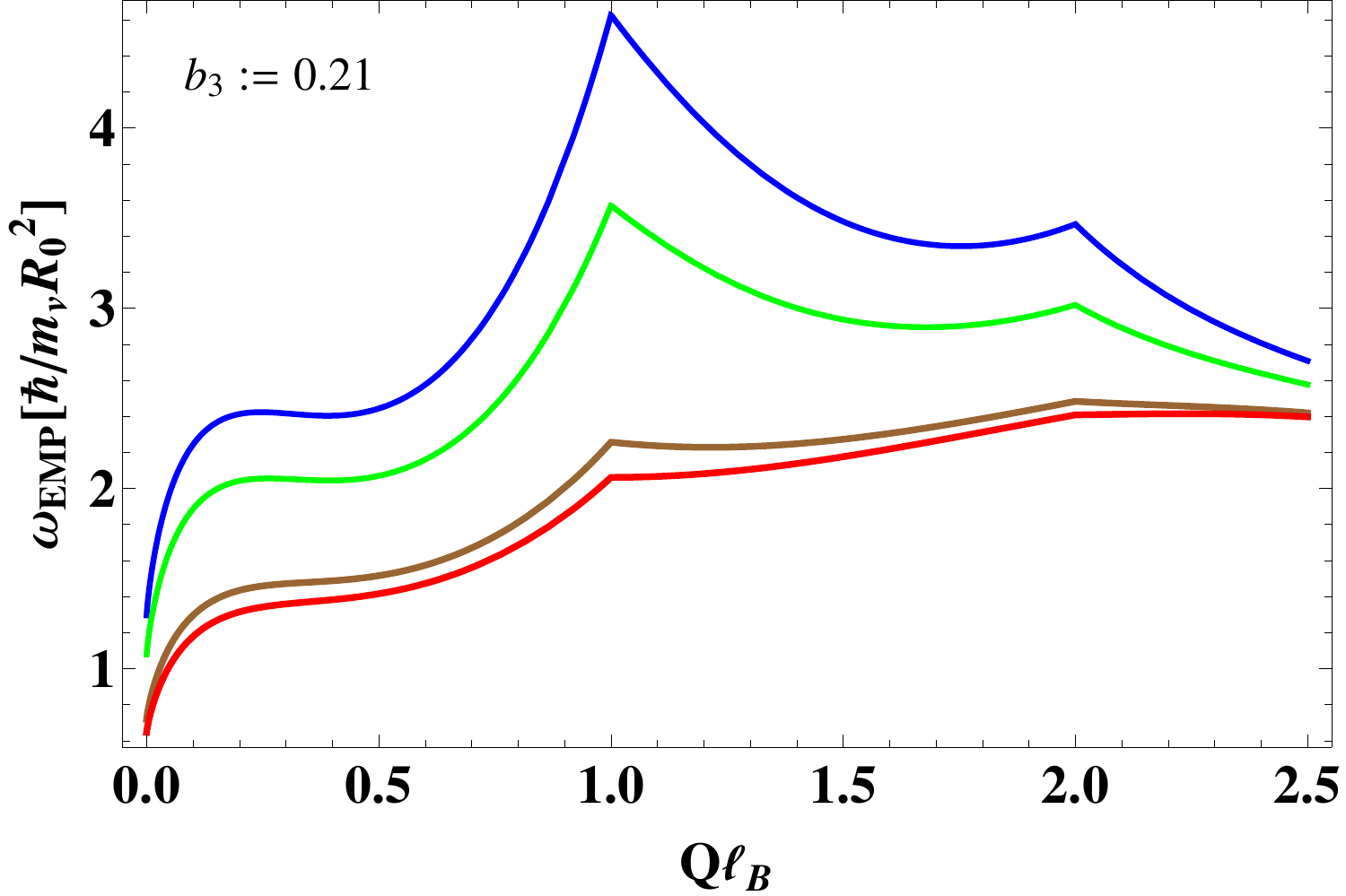}
	\includegraphics[scale=.5]{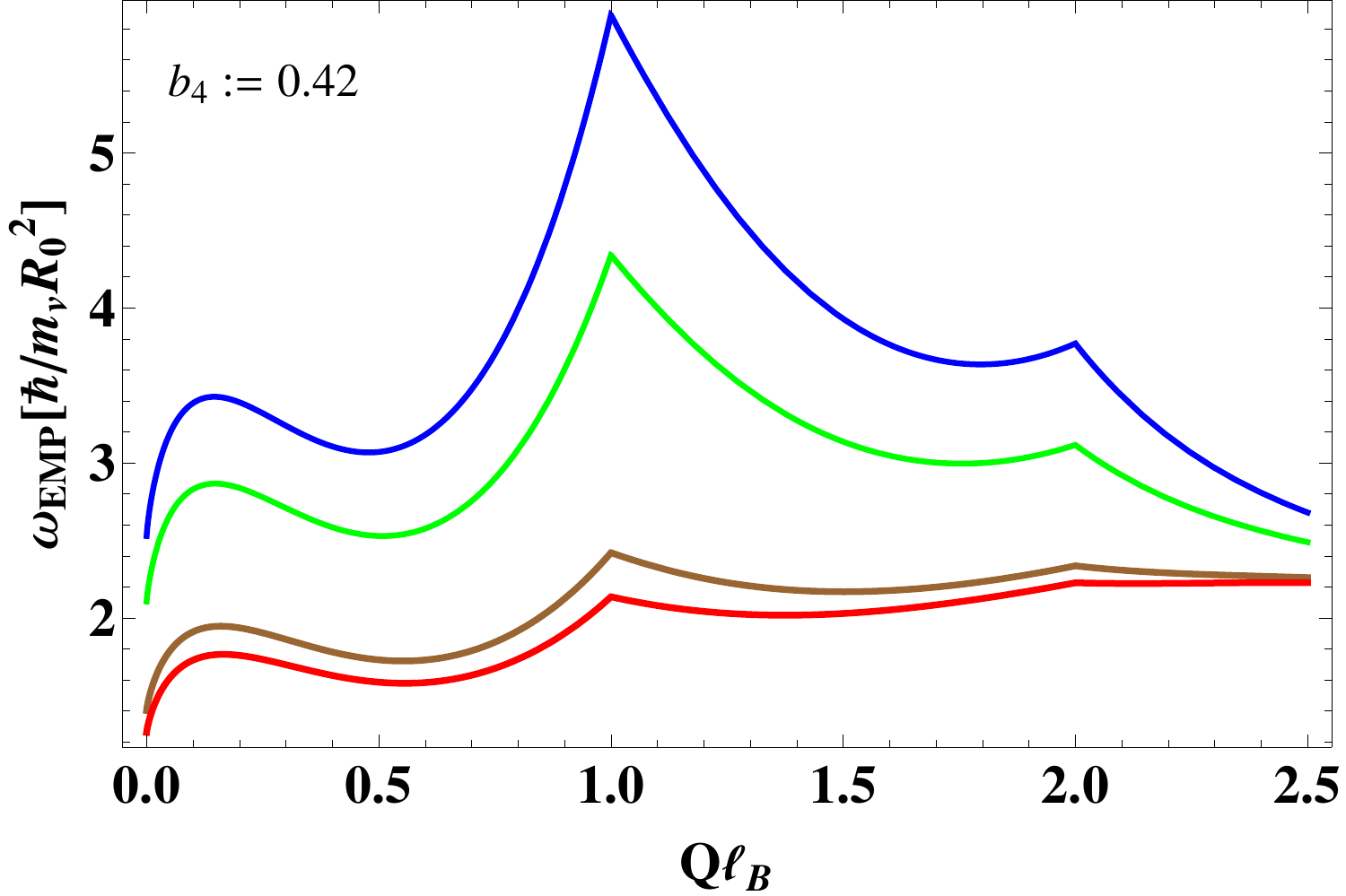}
	\caption{Edge magnetoplasmon dispersion for sharp and smooth edges. The anomalous factor, $\Delta_0/\ell_B$ is varied from $2.0$ to $ 0.5$. Geometry factor: (Left) $b_3=0.21$ and (Right) $b_4=0.42$.}\label{Fig:EMP5}
\end{figure*}

In Figure~\ref{Fig:EMP1}, we plotted Eq.~\ref{Eq:EMPLs} utilizing Eq.~\ref{Eq:EFluxQuan4}. The magnetoplasmon excitation curve is compared with experiments. It is immediately clear that the agreement is favorable. Deviations become significant at high magnetic field and short wavelengths. The high magnetic field regime, where most fractional plateaus show up, are also plotted in Figures~\ref{Fig:EMP3} and~\ref{Fig:EMP5} for both sharp and deformed vortex Hall fluid boundaries, respectively. When magnetic field is increased more vortices are pushed to the boundary forming a secondary boundary near the sample edge. This together with an original fluid boundary creates  a wider double boundary layer, $\Delta_0$. The density profiles and filling factors of both layers differ from each other and from the bulk. Variation in density caused oscillation of charges at boundaries of both layers. Resonant excitations from these layers give rise to the two peaks observed in the plots.  We noticed that an outer edge magnetoplasmons are relatively stronger and sharper compared to a weaker, broader inner edge magnetoplasmons. The resonant excitations at $\ell=1$ and $\ell=2$ with weights $W=3$ and $W=6$ occur when the layers are reasonable separated. The are indicated by the black and red arrows. We particularly note that for both smooth and sharp edges, the second resonant mode becomes well noticed for $\Delta_0 > 2/3\ell_B$. The sharper peak in this limit indicates the more vortices formed and interact to cause excitations. This is in sharp contrast to the $\Delta_0/\ell_B \leq 2/3$ limit, which involves less vortices as seen from the flattening of the curves as boundary deformations become strong.

In conclusion,  we have studied the dynamics of quantized vorticces at the edges of graphene quantum vortex Hall fluid. Edge magneto-excitation frequencies are calculated. We found that for sharp edges under experimental conditions, graphene EMP resonances behave analogously with recent experimental results. For smooth edges, there is gradual distinctions from the sharp edges in the presence of the anomalous term, where an additional EMP peak appear. It becomes well pronounced as the smoothness is increased even though at weaker strengths. We identified the new resonant mode as an Inter-EMP. It originates from  vortices localized and oscillating at second (inner) boundary of a double boundary layer. The present observation brings to light the cause of I-EMP which remained to be detected in graphene experiments. The magnetoplasmon peaks can be probed around $\ell=1$ and $\ell=2$ resonance positions. 

The variation of EMP peaks with boundary deformation means that precise evaluation of $b_M$ is crucial to accurately estimate the actual strength of resonant peaks.


\subsection*{Conflict of interest}
The author(s) declare(s) that there is no conflict of interest regarding the publication of this manuscript.


\end{document}